\documentclass[%
 reprint,
showpacs,preprintnumbers,
 amsmath,amssymb,
 prapplied,
superscriptaddress
]{revtex4-2}

\usepackage{graphicx}
\usepackage{dcolumn}
\usepackage{bm}
\usepackage{physics}
\usepackage{comment}
\usepackage{algpseudocode}
\graphicspath{{./figures/}}

\usepackage{color}

\begin{document}


\title{Semi-classical description of electron dynamics in extended systems under intense laser fields}

\author{Mizuki Tani}
\email{mzktani@atto.t.u-tokyo.ac.jp}
\affiliation{%
 Department of Nuclear Engineering and Management, Graduate School of Engineering, The University of Tokyo,7-3-1 Hongo, Bunkyo-ku, Tokyo 113-8656, Japan
}
\affiliation{%
 Kansai Photon Science Institute, National Institutes for Quantum and Radiological Science and Technology (QST), Kyoto 619-0215, Japan
}%
\author{Tomohito Otobe}%
\email[corresponding author:]{otobe.tomohito@qst.go.jp}
\affiliation{%
 Kansai Photon Science Institute, National Institutes for Quantum and Radiological Science and Technology (QST), Kyoto 619-0215, Japan
}%

\author{Yasushi Shinohara}
\affiliation{%
 Department of Nuclear Engineering and Management, Graduate School of Engineering, The University of Tokyo,7-3-1 Hongo, Bunkyo-ku, Tokyo 113-8656, Japan
}
\affiliation{%
 Photon Science Center, Graduate School of Engineering, The University of Tokyo, 7-3-1 Hongo, Bunkyo-ku, Tokyo 113-8656, Japan
}
\author{Kenichi L. Ishikawa}
\email[corresponding author:]{ishiken@n.t.u-tokyo.ac.jp}
\affiliation{%
 Department of Nuclear Engineering and Management, Graduate School of Engineering, The University of Tokyo,7-3-1 Hongo, Bunkyo-ku, Tokyo 113-8656, Japan
}%
\affiliation{%
 Photon Science Center, Graduate School of Engineering, The University of Tokyo, 7-3-1 Hongo, Bunkyo-ku, Tokyo 113-8656, Japan
}
\affiliation{%
 Research Institute for Photon Science and Laser Technology\, The University of Tokyo, 7-3-1 Hongo, Bunkyo-ku, Tokyo 113-0033, Japan
}




\date{\today}

\begin{abstract}
We propose a semi-classical approach based on the Vlasov equation to describe the time-dependent electronic dynamics in a bulk simple metal under an ultrashort intense laser pulse.
We include in the effective potential not only the ionic Coulomb potential and mean-field electronic Coulomb potential from the one-body electron distribution but also the exchange-correlation potential within the local density approximation (LDA).
The initial ground state is obtained by the Thomas-Fermi
model.
To numerically solve the Vlasov equation, we extend the pseudo-particle method, previously used for nuclei and atomic clusters, to solids, taking the periodic boundary condition into account.
We apply the present implementation to a bulk aluminum (FCC) conventional unit cell irradiated with a short laser pulse.
The optical conductivity, refractive index, extinction coefficient, and reflectivity as well as energy absorption calculated with the Vlasov-LDA method are in excellent agreement with the results by the time-dependent density functional theory and experimental references.
\end{abstract}

\pacs{Valid PACS appear here}
\maketitle


\section{\label{sec:introduction}INTRODUCTION}

The interaction of ultrashort (fs-ps) intense laser pulses with solids is relevant to a wide area of research ranging from high-harmonic generation \cite{Shambhu,Vampa,You,Ndabashimiye,Morimoto} to material machining \cite{Zhigilei,Ilday,Audouard,Betz,Schmidt,Tiinnermann,Mourou,Wang,Gamaly}.
The process of ultrafast laser micromachining, which can suppress heat-affected zone, starts from the energy transfer from the laser to the material by electron excitation, followed by that from the hot electrons to the lattice.
As a result, the material undergoes phase and/or structural transition\cite{Medvedev}, leaving a change of the optical constants or a defect behind \cite{Mazur}, which eventually leads to ablation, drilling, or structuring \cite{Mirza,Silaeva,Rethfeld,Rethfeld2,Rudenko,Chimier,Lorazo,Thorstensen,Kondo,Upadhyay,Ivanov,Ivanov2,Itina2,Garrison,Sakabe,Ishino}.


The comprehensive modeling of laser material machining is highly complex, multi-scale in both time and space, multi-phase (solid, fluid, plasma, cluster, etc.), and possibly accompanied by chemical reactions.
Plasma or continuum models \cite{Zhigilei,Audouard,Matzen,Lehman,Wu,Itina}, for example, have been employed to describe and simulate such processes, advancing understanding.
However, they have difficulties in examining initial transient dynamics before the local thermodynamic equilibrium is reached.

It has now become possible to describe the attosecond-femtosecond electron dynamics under intense laser fields with the time-dependent density-functional theory (TDDFT) \cite{salmon,Yabana,Magyar,Otobe} or time-dependent density-matrix methods \cite{DM,Hirori,Sanari}.
TDDFT is an {\it ab initio} method that offers a good compromise between accuracy and computational feasibility.
Its computational cost is, however, still very high, especially, if one wants to perform long-timescale simulation, coupling it with molecular dynamics and electromagnetic-field analysis.

In TDDFT, each electron orbital satisfies the time-dependent Kohn-Sham (TDKS) equation [see Eq.~(\ref{TDKS}) below].
The leading order of a semiclassical $\hbar$ expansion of the TDKS equation reduces to the Vlasov equation, which describes the temporal evolution of the electron distribution function in phase space.
Thus, Vlasov-based approaches are expected to be a cost-effective alternative to TDDFT, in particular, for metals.
Such approaches have previously been applied to ionization and explosion dynamics of molecules \cite{Ishikawa,c60} and metal clusters \cite{Giglio,Fennel,Kohn,Plagne,Domps,Domps2}.
The Vlasov equation is numerically solved with so-called pseudo-particle methods in these studies, which represent the electron cloud as an assembly of classical test particles whose motion is governed by Newton’s equations of motion.
There are several reports of application to Na clusters, well agreeing with TDDFT results \cite{Plagne,Domps,Domps2,Calvayrac,Feret}.

In this paper, we extend the pseudo-particle method based on the Vlasov equation to the description of electron dynamics in extended systems under intense laser fields. 
The effective potential acting on the electrons contains not only the ionic potential, interelectronic Hartree potential, and interaction with laser but also the exchange-correlation potential within the local-density approximation (LDA), and incorporates the periodic boundary condition. 
We apply the present method to bulk aluminum.
The calculated optical conductivity, refractive index, extinction coefficient, and reflectivity as well as energy absorption are in excellent agreement with TDDFT calculations and experimental references.

The present paper is organized as follows. Section \ref{sec:methods} describes our simulation methods. 
We review the Vlasov equation and describe our numerical implementations with the periodic boundary condition. 
In Sec.~\ref{sec:results} we describe numerical application to bulk aluminum and compare the results with TDDFT and measurement values. 
The conclusions are given in Sec.~\ref{sec:conclusions}.

\section{\label{sec:methods}METHODS}
\subsection{\label{subsec:Vlasov equation}Vlasov equation}
Among the methods for treating quantum many-body dynamics, TDDFT provides a feasible computational framework for treating electronic systems' optical response or charged particles' collision phenomena \cite{NaAr}. 
The time propagation of a $N_e$-electron system comes down to solving a set of equations for the Kohn-Sham orbitals $\{\phi_i(\mathbf{r},t)\}$ that evolve in a self-consistent mean field \cite{TDDFT},
\begin{equation}
    \mathrm{i}\hbar \frac{\partial}{\partial t}\phi_i(\mathbf{r},t) = h_{\mathrm{KS}}[n_{\mathrm{e}}(\mathbf{r},t)]\phi_i(\mathbf{r},t), \label{TDKS}
\end{equation}
where,
\begin{equation}
    h_{\mathrm{KS}}[n_{\mathrm{e}}(\mathbf{r},t)] = -\frac{\hbar^2}{2m}\nabla^2 + V_{\mathrm{eff}}[n_{\mathrm{e}}(\mathbf{r},t)],
\end{equation}
denotes the Kohn-Sham Hamiltonian, $m$ the electron mass, $V_{\rm eff}$ the effective potential (see below), and 
the time-dependent electron density $n_e({\bf r},t)$ is defined as,
\begin{equation}
    n_{\mathrm{e}}(\mathbf{r},t) = \sum_{i=1}^{N_e} |\phi_i(\mathbf{r},t)|^2.
\end{equation}
Analogously, the density operator $\hat{\rho}(t)$ is defined as,
\begin{equation}
    \mel{\mathbf{r}}{\hat{\rho}(t)}{\mathbf{r}'} = \sum_{i=1}^{N_e} \phi_i^*(\mathbf{r},t)\phi_i(\mathbf{r}',t),
\end{equation}
whose evolution is govenrned by the von-Neumann equation (vNE),
\begin{equation}
    \frac{\partial}{\partial t}\hat{\rho}(t) = -\frac{\mathrm{i}}{\hbar}\left[ \hat{h}_{\mathrm{KS}}(t), \hat{\rho}(t) \right]. \label{TDVN}
\end{equation}

Performing the Wigner transformation \cite{Wigner} and taking the limit $\hbar \to 0$, the density operator $\hat{\rho}(t)$ is mapped onto a real function $f(\mathbf{r},\mathrm{\mathbf{p}},t)$, which obeys the Vlasov equation,
\begin{align}
    \frac{\partial}{\partial t}&f(\mathbf{r},\mathrm{\mathbf{p}},t)\notag\\
    &= -\frac{\mathrm{\mathbf{p}}}{m}\cdot \nabla_{\mathbf{r}}f(\mathbf{r},\mathrm{\mathbf{p}},t)+\nabla_{\mathbf{r}}V_{\mathrm{eff}}[n_{\mathrm{e}}(\mathbf{r},t)]\cdot \nabla_{\mathrm{\mathbf{p}}}f(\mathbf{r},\mathrm{\mathbf{p}},t), \label{Vlasov}
\end{align}
which is a classical alternative to the vNE, Eq.~(\ref{TDVN}), where $\mathrm{\mathbf{p}}$ is the electron canonical momentum. Here, 
$f(\mathbf{r},\mathbf{p},t)$ is interpreted as the electron distribution in phase space.

The effective potential $V_{\mathrm{eff}}$ is a functional of the electron density distribution $n_e({\bf r},t)$ and decomposed into,
\begin{equation}
    V_{\mathrm{eff}}[n_{\mathrm{e}}(\mathbf{r},t)] = V_{\mathrm{Coulomb}}[n_{\mathrm{e}}(\mathbf{r},t)] + V_{\mathrm{xc}}[n_{\mathrm{e}}(\mathbf{r},t)] + V_{\mathrm{ext}}(\mathbf{r},t), \label{Veff}
\end{equation}
with the exchange-correlation potential $V_{xc}$, external field potential $V_{\rm ext}$, and, 
\begin{equation}
    V_{\mathrm{Coulomb}}[n_{\mathrm{e}}(\mathbf{r},t)] = \sum_{i}V_{\mathrm{ps}}(\mathbf{r}-\mathbf{r}_i) + V_{\mathrm{\mathrm{H}}}[n_{\mathrm{e}}(\mathbf{r},t)],
\end{equation}
where $i$, $V_{\mathrm{ps}}$ and $V_{\mathrm{H}}$ denote the label of ions and the spherically symmetric ionic pseudopotential and the electron-electron Hartree potential, respectively. 

Several previous works for Na clusters have used their original pseudo potentials \cite{Fennel,Giglio}, adjusted so that the simulation results reproduce the static and dynamical properties of the system. In this work, instead, we employ the modified Heine-Abarenkov type local pseudo potential for $V_{\mathrm{ps}}$,
\begin{equation}
    V_{\mathrm{ps}}(r) = -\frac{z}{R}e\left\{ \frac{1}{r} \left[ 1-(1+\beta r)\mathrm{e}^{-\alpha r} - A\mathrm{e}^{-r} \right] \right\} (r=|\mathbf{r}|),
\end{equation}
where $z$ is the number of the valence electrons, and $A, R, \alpha$, and $\beta$ are material dependent parameters determined by \it ab initio \rm density functional formalism in Ref.~\cite{Vps}, thus, independently from Vlasov simulations. Their values for the bulk aluminum crystal are $A=3.574 \ \mathrm{a.u.}, \alpha=3.635 \ \mathrm{a.u.}, \beta=0.8343 \ \mathrm{a.u.}, R=0.334 \ \mathrm{a.u.}, z=3$.
$V_{\mathrm{H}}$ is evaluated by solving the Poisson equation,
\begin{equation}
    \label{eq:Poisson}
    \Delta V_{\mathrm{H}} [n_e(\mathbf{r},t)] = -4\pi e n_e(\mathbf{r},t).
\end{equation}

Here, let us introduce a real-space simulation box $\Omega$, on which the periodic boundary condition is imposed, and translation vectors $\mathbf{G}$. $\Omega$ is defined as,
\begin{equation}
    \Omega = \left\{ \mathbf{r} = \sum_{j=x,y,z} a_j\mathbf{e}_j \Bigg{|} \ 0\le a_j<1 \right\},
\end{equation}
where $\{\mathbf{e}_j\}$ are the lattice vectors along the $j$-axis ($j=x, y, z$), whose lengths are denoted by $L_j = |\mathbf{e}_j|$. Integrals with respect to $\mathbf{r}$ are taken over $\Omega$ in what follows. The translation vectors are given by,
\begin{equation}
    \mathbf{G} = \sum_{j=x,y,z} M_j\mathbf{e}_j \ (M_j = 0, \pm1, \pm2, \cdots).
\end{equation}

Taking the periodic boundary condition into account, the Coulomb terms $V_{\mathrm{ps}}$ and $V_{\mathrm{\mathrm{H}}}$ are represented as a Fourier series expansion.
The pseudo potential term is rewritten as,
\begin{gather}
   \sum_{i}^{\infty}V_{\mathrm{ps}}(\mathbf{r}-\mathbf{r}_i)=\sum_{\mathbf{G},i=1}^{N_{\mathrm{ion}}}V_{\mathrm{ps}}(\mathbf{r}-\mathbf{r}_i-\mathbf{G}),
\end{gather}
where $N_{\mathrm{ion}}$ denotes the number of ions in $\Omega$ and,
\begin{gather}
    \sum_{\mathbf{G}}V_{\mathrm{ps}}(\mathbf{r}-\mathbf{r}_i-\mathbf{G})\qquad \qquad \qquad \qquad \qquad \notag \\
    \qquad = \mathcal{F}^{-1} \left[ \sum_{i} \mathrm{e}^{-\mathbf{Q}\cdot\mathbf{r}_i}\left\{ V_{\mathrm{ps}}(Q) +\frac{4\pi}{Q}z \right\} \right]. \label{Vps}
\end{gather}
with $\mathbf{Q}$ being the coordinates in the Fourier domain ($Q=|\mathbf{Q}|$), $\mathcal{F}[\cdot]$ the Fourier series expansion within $\Omega$, and,
\begin{gather}
    V_{\mathrm{ps}}(Q) = 4\pi z e R^2 \left[ -\frac{1}{\left( QR \right)^2} + \frac{1}{\left( QR \right)^2 + \alpha^2} \right.  \notag\\ \qquad \left. + \frac{2\alpha \beta}{\left\{ \left( QR \right)^2 + \alpha^2 \right\}^2} + \frac{2A}{\left\{ \left( QR \right)^2 + 1 \right\}^2}
    \right],
\end{gather}
One obtains the solution of the Poisson equation (Eq.~(\ref{eq:Poisson})) for the electron density $n_{\mathrm{e}}$ given within $\Omega$ as,
\begin{equation}
    V_{\mathrm{\mathrm{H}}}[n_{\mathrm{e}}(\mathbf{r},t)] = \mathcal{F}^{-1} \left[ \mathcal{F} \left[ n_{\mathrm{e}}(\mathbf{r},t) \right] \frac{4\pi e}{Q^2} \right].
\end{equation}

For the exchange-correlation potential $V_{\mathrm{xc}}$, one employs the LDA by Perdew and Zunger \cite{PZ}.
The laser-electron interaction is described in the length gauge,
\begin{equation}
    V_{\mathrm{ext}}(\mathbf{r},t) = -e\mathbf{E}(t)\cdot \mathbf{r},
\end{equation}
within the dipole approximation, where ${\bf E}$ denotes the laser electric field vector.
In this case, ${\bf p}$ becomes the kinetic momentum, and, thus, the electronic current density $\mathbf{J}(t)$ averaged over $\Omega$ is given by,
\begin{equation}
    \mathbf{J}(t) = \frac{1}{|\Omega|}\iint_{\Omega} \left(-e \frac{\mathbf{p}}{m}\right) f(\mathbf{r},\mathbf{p},t) \dd \mathbf{r} \dd \mathbf{p}.
    \label{eq:current density}
\end{equation}

\subsection{\label{subsec:numerical implementations} Numerical implementations}
\subsubsection{\label{subsubsec:pseudo-particle method} Pseudo-particle method}
The direct propagation of the distribution function would require the treatment of six-dimensional time-dependent function on grids \cite{gridVlasov}. To avoid such a massive computation, one introduces the pseudo-particle method \cite{Bertsch,Giglio,Fennel,Kohn}, where the distribution function $f(\mathbf{r},\mathbf{p},t)$ is expressed by a set of pseudo-particles with mass $m$ as,
\begin{equation}
    f(\mathbf{r},\mathbf{p},t) = \frac{1}{N_s}\sum_{i=1}^{N_{\mathrm{pp}}}g_r \left( \mathbf{r}-\mathbf{r}_i(t) \right) g_p \left( \mathbf{p}-\mathbf{p}_i(t) \right). \label{ftp}
\end{equation}
Here $\mathbf{r}_i, \mathbf{p}_i$ are the position and canonical momentum of each pseudo particle labeled by $i$. The total number of pseudo particles $N_{pp}$ is given by $N_{pp}=N_sN_e$, where $N_{s}$ and $N_e$ are the number of pseudo particles per electron and the total number of the electrons contained in $\Omega$, respectively. 
Statistical error is reduced by increasing $N_s$. $N_s$ is set to 10000 in this study.
$g_r(\mathbf{r})$ and $g_p(\mathbf{p})$ denote smoothing kernel functions for the position and momentum, respectively, of Gaussian forms,
\begin{align}
    g_r (\mathbf{r}) &= \sum_{\{\mathbf{G}\}} \frac{1}{\pi^{3/2}d_r^3}\exp(-|\mathbf{r} + \mathbf{G}|^2/d_r^2), \label{eq:smoothing-kernel-r}\\
    g_p (\mathbf{p}) &= \frac{1}{\pi^{3/2}d_p^3}\exp(-|\mathbf{p}|^2/d_p^2),
\end{align}
where $d_r$ and $d_p$ are smoothing widths. Only the nearest neighbor cells are included in summation over $\mathbf{G}$ in Eq.~\eqref{eq:smoothing-kernel-r}. 
The kernel functions are normalized as,
\begin{align}
    \int_{\Omega} g_r (\mathbf{r}) \dd \mathbf{r} &= 1, \\
    \int g_p (\mathbf{p}) \dd \mathbf{p} &= 1,
\end{align}
so that,
\begin{gather}
    \iint_{\Omega} f({\bf r},{\bf p},t)\, \dd \mathbf{r} \dd \mathbf{p} = N_e.
\end{gather}
The scattering cross-section of the electron and the effective potential is adjusted through $d_r$; the smaller $d_r$, the larger the cross-section. Here we use $d_r = 0.575 \ \mathrm{a.u.}$ so that the time-dependent energy absorption reproduces the TDLDA results. 
In the present collision-less case, $d_p$ is not used explicitly. 

The field quantities such as $V_{\mathrm{eff}}$ and $n_e$ are evaluated on three-dimensional grids discretized into $N_j$ ($j=x,y,z$) intervals on the $j$ axis with a spatial step $\Delta j = L_j/N_j$. In our calculation we set $\Delta x = \Delta y = \Delta z = 0.5 \ \mathrm{a.u.}$ Here, $d_r/\Delta x\simeq 1.15$ is a good parameterization leading to stable simulation \cite{Fennel}. It should be noticed that $d_r$ is the only adjustable parameter in our formalism.
The electron density on a grid point $\mathbf{r}$ is calculated as,
\begin{equation}
    n_e(\mathbf{r},t) = \int \dd \mathbf{p} f(\mathbf{r},\mathbf{p},t)
    = \frac{1}{N_s}\sum_{i=1}^{N_{\mathrm{pp}}}g_r \left( \mathbf{r}_i(t)-\mathbf{r} \right). \label{n_e}
\end{equation}
The current density $\mathbf{J}(t)$ [Eq.~\eqref{eq:current density}] is evaluated as,
\begin{equation}
    \mathbf{J}(t) = -\frac{1}{|\Omega|} \frac{e}{N_s}\sum_{i=1}^{N_{\mathrm{pp}}} \frac{\mathbf{p}_i(t)}{m}.
\end{equation}

The Hamiltonian in pseudo-particle representation is written as,
\begin{equation}
    H_{\mathrm{pp}} = \frac{1}{N_s}\sum_i^{N_{\mathrm{pp}}} \left[ \frac{\mathbf{p}^2_i(t)}{2m} + \int_{\Omega} V_{\mathrm{eff}}(\mathbf{r},t)g_r \left( \mathbf{r}_i-\mathbf{r} \right) \dd \mathbf{r} \right]. \label{ppH}
\end{equation}
The motion of each pseudo particle is governed by the Newton equations under the effective potential $V_{\mathrm{eff}}$ with the periodic boundary condition as,
\begin{equation}
    \dot{\mathbf{r}_i}=\frac{\mathbf{p}_i}{m}, \ \dot{\mathbf{p}}_i = -\int_{\Omega} V_{\mathrm{eff}}(\mathbf{r})\nabla_{\mathbf{r}_i}  g_r(\mathbf{r}_i-\mathbf{r})\mathrm{d}\mathbf{r}. \label{Newton}
\end{equation}
As long as pseudo-particle canonical variables $\mathbf{r}_i, \mathbf{p}_i$ obey the Newton equation (Eq.~(\ref{Newton})), one-body distribution (Eq.~(\ref{ftp})) satisfies the Vlasov equation (Eq.~(\ref{Vlasov})).
 The force term is given as the gradient of the $N_{\mathrm{pp}}$-body Hamiltonian $H_{\mathrm{pp}}$. 
 One numerically integrates it as,
\begin{align}
    \int V_{\mathrm{eff}}(\mathbf{r})&\nabla_{\mathbf{r}}  g_r(\mathbf{r}_i-\mathbf{r})\mathrm{d}\mathbf{r} \notag\\
    &\simeq \sum_{\mathbf{r}\in \Omega} V_{\mathrm{eff}}(\mathbf{r})\nabla_{\mathbf{r}_i}  g_r(\mathbf{r}_i-\mathbf{r})\Delta x\Delta y\Delta z,
\end{align}
using the analytical form of $\nabla_{\mathbf{r}_i}  g_r(\mathbf{r}_i-\mathbf{r})$,
\begin{align}
        \nabla_{\mathbf{r}_i} &g_r (\mathbf{r}_i-\mathbf{r}) \notag \\
    &= \sum_{\{\mathbf{G}\}} \frac{-2(\mathbf{r}_i - \mathbf{r} + \mathbf{G})}{\pi^{3/2}d_r^5}\exp(-|\mathbf{r}_i - \mathbf{r} + \mathbf{G}|^2/d_r^2).
\end{align}
The integration of Eq.~(\ref{Newton}) is performed by the Verlet method \cite{verlet}
with time step $\Delta t = 0.02 \ \mathrm{a.u.}$ Particles exiting $\Omega$ are to re-enter $\Omega$ from the other side.


\subsubsection{\label{subsubsec:ground state}Ground state}

The initial state is the stationary solution of the Vlasov equation described by the Thomas-Fermi model. The total energy functional,
\begin{gather}
    E_{\mathrm{all}}[n_{\mathrm{e}}(\mathbf{r})] = \int_{\Omega} \Bigl[ \frac{3}{10}\frac{\hbar^2(3\pi^2)^{\frac{2}{3}}}{m}n_{\mathrm{e}}^{\frac{5}{3}}(\mathbf{r}) + \frac{1}{2}V_{\mathrm{H}}(\mathbf{r})n_{\mathrm{e}}(\mathbf{r}) \notag \\ \quad
    + \sum_{\mathbf{G},i=1}^{N_{\mathrm{ion}}}V_{\mathrm{ps}}(\mathbf{r}-\mathbf{r}_i-\mathbf{G})n_{\mathrm{e}}(\mathbf{r}) 
    + E_{\mathrm{xc}}[n_{\mathrm{e}}(\mathbf{r})] \Bigr] \mathrm{d} \mathbf{r},
\end{gather}
is variationally minimized with respect to $n_{\mathrm{e}}(\mathbf{r})$ under the constraint that the box $\Omega$ contains $N_e$ electrons.
This leads to the following coupled equations,
\begin{equation}
    \frac{\hbar^2}{2m}\left[ 3\pi^2n_e(\mathbf{r}) \right]^{2/3} + V_{\mathrm{eff}}(\mathbf{r}) = \mu, \label{tf}
\end{equation}
\begin{equation}
    V_{\mathrm{eff}}(\mathbf{r}) = V_{\mathrm{Coulomb}}[n_{\mathrm{e}}(\mathbf{r})] + V_{\mathrm{xc}}[n_{\mathrm{e}}(\mathbf{r})], \label{tf_veff}
\end{equation}
where $\mu$ denotes the chemical potential, playing the role of a Lagrange multiplier.
These equations are to be solved for $n_e({\bf r})$ self-consistently.

An adopted algorithm to solve the coupled equations Eq.~(\ref{tf}) and Eq.~(\ref{tf_veff}) is shown in Fig.~\ref{GSscheme}.
\begin{figure}[tb]
\begin{algorithmic}[1]
    \Procedure{Ground state preparation}{}
        \State{(--Initialization--)}
        \State{initial guess of $\mu$ and $n_e^{\mathrm{in}}$}
        \State{ }
        \State{(--Self-consistent determination of $\mu$ and $n_e$--)}
        \While{$\Delta n> \epsilon \, (\epsilon=10^{-7})$}
            \State{set pseudo-particle position \{$\mathbf{r}_i$\}}
            \State{$\{\mathbf{r}_i\}\mapsto n_{\mathrm{ps}}(\mathbf{r})$ using Eq.~(\ref{n_e})}
            \State{$n_{\mathrm{ps}} \mapsto V_{\mathrm{eff}}[n_{\mathrm{ps}}](\mathbf{r})$ using Eq.~(\ref{tf_veff})}
            \State{$V_{\mathrm{eff}}[n_{\mathrm{ps}}](\mathbf{r}) \mapsto n_e$}
            \State{find appropriate $\mu$}
            \State{$\Delta n = \int_{\Omega} \dd \mathbf{r}|n_e^{\mathrm{in}}(\mathbf{r})-n_e(\mathbf{r})|$}
            \State{$n_e^{\mathrm{in}}=n_e$}
        \EndWhile
        \State{ }
        \State{(--Set Pseudo-particle Momenta--)}
        \For{$i=1,N_{pp}$}
            \While{$p>p_f$}
                \State{random number $p_x$ ($0 \le |p_x| \le p_f$)}
                \State{random number $p_y$ ($0 \le |p_y| \le p_f$)}
                \State{random number $p_z$ ($0 \le |p_z| \le p_f$)}
                \State{$p=\sqrt{p_x^2+p_y^2+p_z^2}$}
            \EndWhile
            \State{$\mathbf{p}_i=(p_x,p_y,p_z)$}
        \EndFor
    \EndProcedure
\end{algorithmic}
\caption[GS algorithm]{Algorithm for the ground state preparation}
\label{GSscheme}
\end{figure}

First, the chemical potential $\mu$ and the electron density $n_e^{\mathrm{in}}$ in the real space are guessed so that the total number of electrons within $\Omega$ is $N_e$ (line 3). 
Then, one distributes pseudo particles according to the guessed $n_e^{\mathrm{in}}$ using random numbers (line 7, also see below). 
The electron density $n_{\mathrm{ps}}$ realized by the pseudo-particle distribution is calculated through Eq.~(\ref{n_e}) (line 8). 
The effective potential $V_{\mathrm{eff}}({\bf r})$ is obtained by substituting $n_{\mathrm{ps}}$ into the right-hand side of Eq.~ (\ref{tf_veff})  (line 9). 
Then, we update the electron density $n_e({\bf r}) $ by substituting thus obtained $V_{\mathrm{eff}}({\bf r})$ to the Eq.~(\ref{tf}) and solving it with respect to $n_e$ (line 10), simultaneously updating $\mu$ by a bisection method to satisfy the condition that the total number of electrons is $N_e$ (line 11).
The updated $n_e$ is used as $n_e^{\mathrm{in}}$ in the next iteration of the loop (line 13).
One repeats the above operations till convergence, $\int_{\Omega} \dd \mathbf{r}|n_e^{\mathrm{in}}-n_e|<\epsilon$, where we set $\epsilon=10^{-7}$ here for crystalline Al (line 12). 
After convergence, one distributes the momenta of the pseudo particles uniformly within the local Fermi radius $p_f$ by acceptance-rejection sampling of uniform pseudo-random numbers (lines 17-25).

The algorithm to distribute the pseudo particles (line 7 in Fig.~\ref{GSscheme}) is shown in Fig.~\ref{dist}. 
\begin{figure}[tb]
\begin{algorithmic}[1]
\Procedure{how to set pseudo particle position}{}
    \State{(--\# of Pseudo Particles around $\mathbf{r}_s$--)}
    \State{a sub-grid point $\mathbf{r}_s=(x_s,y_s,z_s)$}
    \State{calculate $n_e^{\mathrm{inp}}(\mathbf{r}_s)$ by interpolation}
    \State{the number of pseudo particles $n_e^{\mathrm{inp}} \mapsto N_{\mathrm{pp}}^{\mathrm{local}}$}
    \State{ }
    \State{(--Distribute Pseudo Particles around $\mathbf{r}_s$--)}
    \For{$l=1,N_{\mathrm{pp}}^{\mathrm{local}}$}
        \State{random number $R_x$ ($-0.5\le R_x \le0.5$)}
        \State{random number $R_y$ ($-0.5\le R_y \le0.5$)}
        \State{random number $R_z$ ($-0.5\le R_z \le0.5$)}
        \State{$x_l=x_s+R_xL_x/N_{\mathrm{inp}}^x$}
        \State{$y_l=y_s+R_yL_y/N_{\mathrm{inp}}^y$}
        \State{$z_l=z_s+R_zL_z/N_{\mathrm{inp}}^z$}
        \State{$l$-th pseudo particle position is}
        \State{$\mathbf{r}_l=(x_l,y_l,z_l)$}
    \EndFor
\EndProcedure
\end{algorithmic}
\caption[distribution algorithm]{Algorithm for pseudo particle distribution}
\label{dist}
\end{figure}

We introduce sub-grid points (line 3) by dividing each voxel of the computational grid into $N_{\mathrm{inp}}^j$ regions along the $j$-axis ($j=x,y,z$). 
The electron density $n_e^{\mathrm{inp}}$ on each sub-grid point is evaluated, based on the trilinear interpolation from those of the surrounding eight computational grid points, from which one calculates the number of pseudo particles $N_{\mathrm{pp}}^{\mathrm{local}}$ around the sub-grid point (lines 4-5).
Then, the $N_{\mathrm{pp}}^{\mathrm{local}}$ pseudo particles are uniformly distributed around the sub-grid point using random numbers (lines 8-17). 

\subsection{\label{subsec:linear response evaluation} Linear response}

We evaluate the linear optical response via impulse response by doing dynamical simulations with the initial pseudo-particle momenta $\mathbf{p}_i$ shifted from the ground-state values $\mathbf{p}_i^{\mathrm{GS}}$ by a small amount $\Delta \mathbf{p}$,
\begin{equation}
    \mathbf{p}_i = \mathbf{p}_i^{\mathrm{GS}} + \Delta \mathbf{p},
\end{equation}
where $\Delta \mathbf{p} = (0, 0, 0.1 \ \mathrm{a.u.})$ in this study.
This is equivalent to the application of an impulse electric field,
\begin{equation}
    \mathbf{E}(t) = -\frac{1}{e}\Delta \mathbf{p}\,\delta(t),
\end{equation}
where $\delta(t)$ is the delta function.
Noting that this field has a constant power spectrum across all frequencies, one can readily obtain the optical conductivity as,
\begin{equation}
    \sigma_{mn}(\omega) = -\frac{e\hat{J}_m(\omega)}{\Delta p_n}\quad (m,n = x,y,z),
\end{equation}
where $\Delta p_m$ and $\hat{J}_m$ ($m,n = x,y,z$) denote the $m$ component of the momentum shift and the temporal Fourier transform of the current density, respectively. The fast Fourier transformation algorithm \cite{FFT} is used for the evaluation of $\hat{J}_m(\omega)$.
Assuming isotropic media, the dielectric function $\varepsilon_{mm}(\omega)$, the complex refractive index $n(\omega)$, and the reflectivity $R(\omega)$ are given by,
\begin{align}
    \varepsilon_{mm}(\omega) &= 1+4\pi\mathrm{i}\frac{\sigma_{mm}(\omega)}{\omega},\\
    n(\omega) &= \sqrt{\varepsilon_{mm}(\omega)},\\
    R(\omega) &= \left| \frac{\sqrt{\varepsilon_{mm}(\omega)}-1}{\sqrt{\varepsilon_{mm}(\omega)}+1} \right|^2,\\
    \ \notag
\end{align}
respectively, especially, $\varepsilon_{xx}(\omega)=\varepsilon_{yy}(\omega)=\varepsilon_{zz}(\omega)$.

\section{\label{sec:results} RESULTS}

In this section, we compare the results of the Vlasov-LDA simulations for extended systems described in the previous section with the experimental values and the TDDFT results obtained by the open source code SALMON \cite{salmon,salmon2,salmon3}.
We take aluminum as a target material. For Vlasov-LDA, simulation parameters are $N_s=10000$, $N_e=12$, and time step $\Delta t=0.025 \ \mathrm{a.u.}$. For TDDFT, we employ a norm-conserving pseudopotential \cite{FHI} and the LDA functional \cite{PZ}, with the number of k-points $48^3$, number of real-space grids $14^3$, and $dt=0.15 \ \mathrm{a.u}$. We assume an external electric field linearly polarized along the $\Gamma-X$ direction of the following temporal profile: 
\begin{equation}
    E(t)=E_0\sin\left[\omega\left(t-\frac{T}{2}\right)\right]\sin^2\left(\frac{t}{T}\pi\right) \ (0\le t\le T),
\end{equation}
where $E_0$ denotes the field amplitude, $\hbar \omega$ the photon energy, and $T$ the (foot-to-foot) full pulse duration. The corresponding full
width at half maximum duration of the laser intensity profile
is about $0.36T$. 

\subsection{\label{subsec:linear response} Linear response}

\begin{figure}[tb]
  \centering
  \includegraphics[keepaspectratio,width=\hsize]{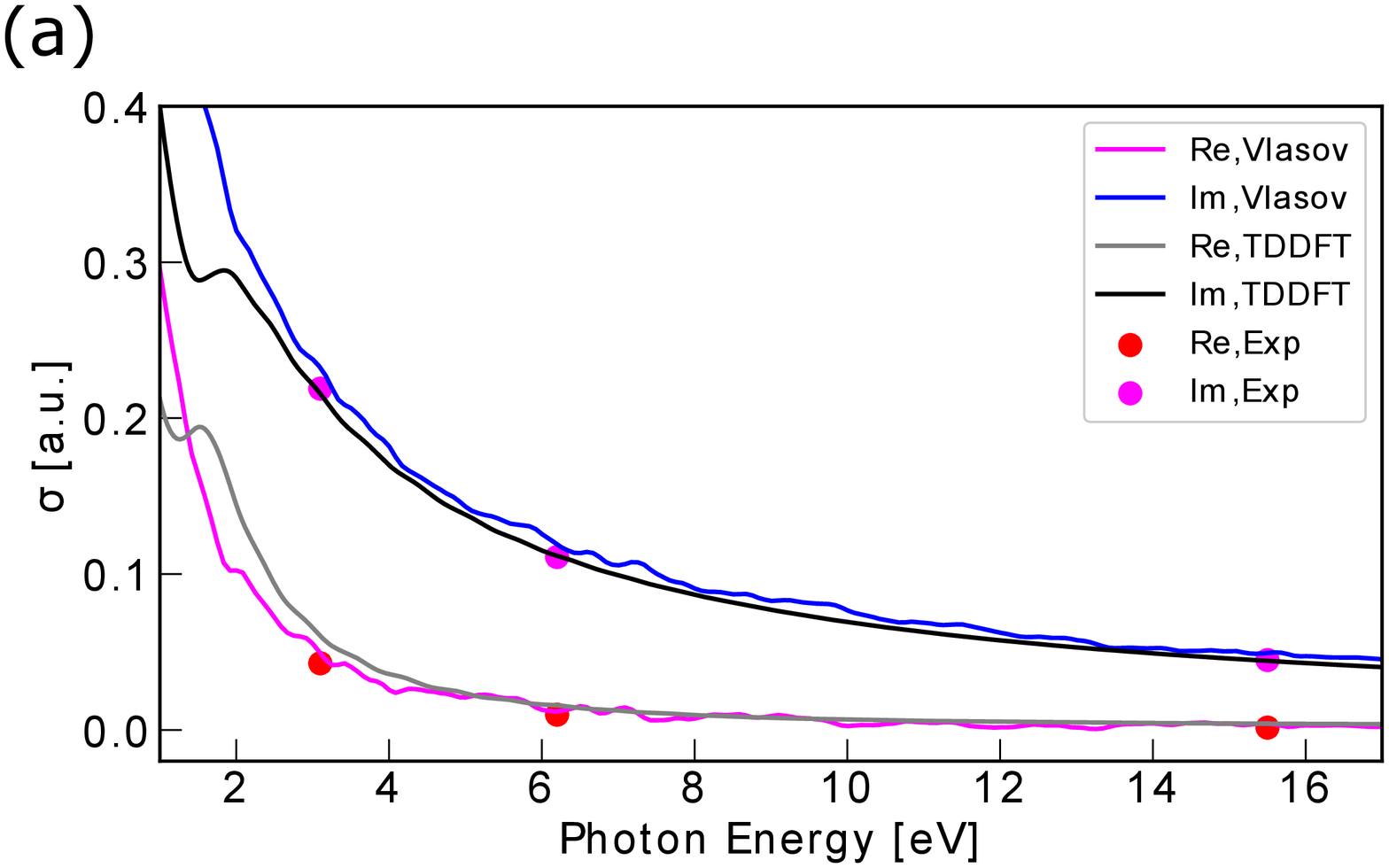}
  \includegraphics[keepaspectratio,width=\hsize]{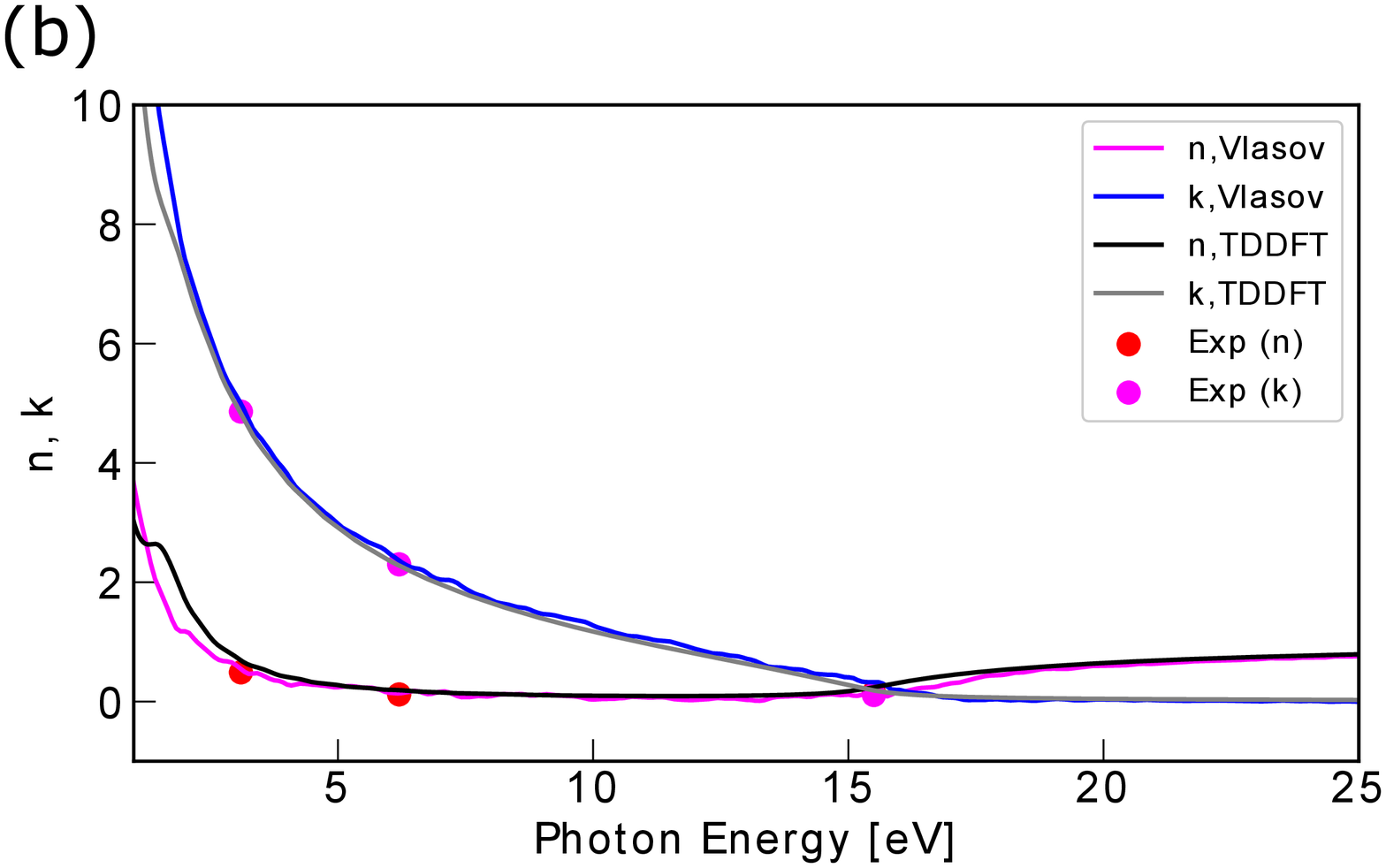}
  \includegraphics[keepaspectratio,width=\hsize]{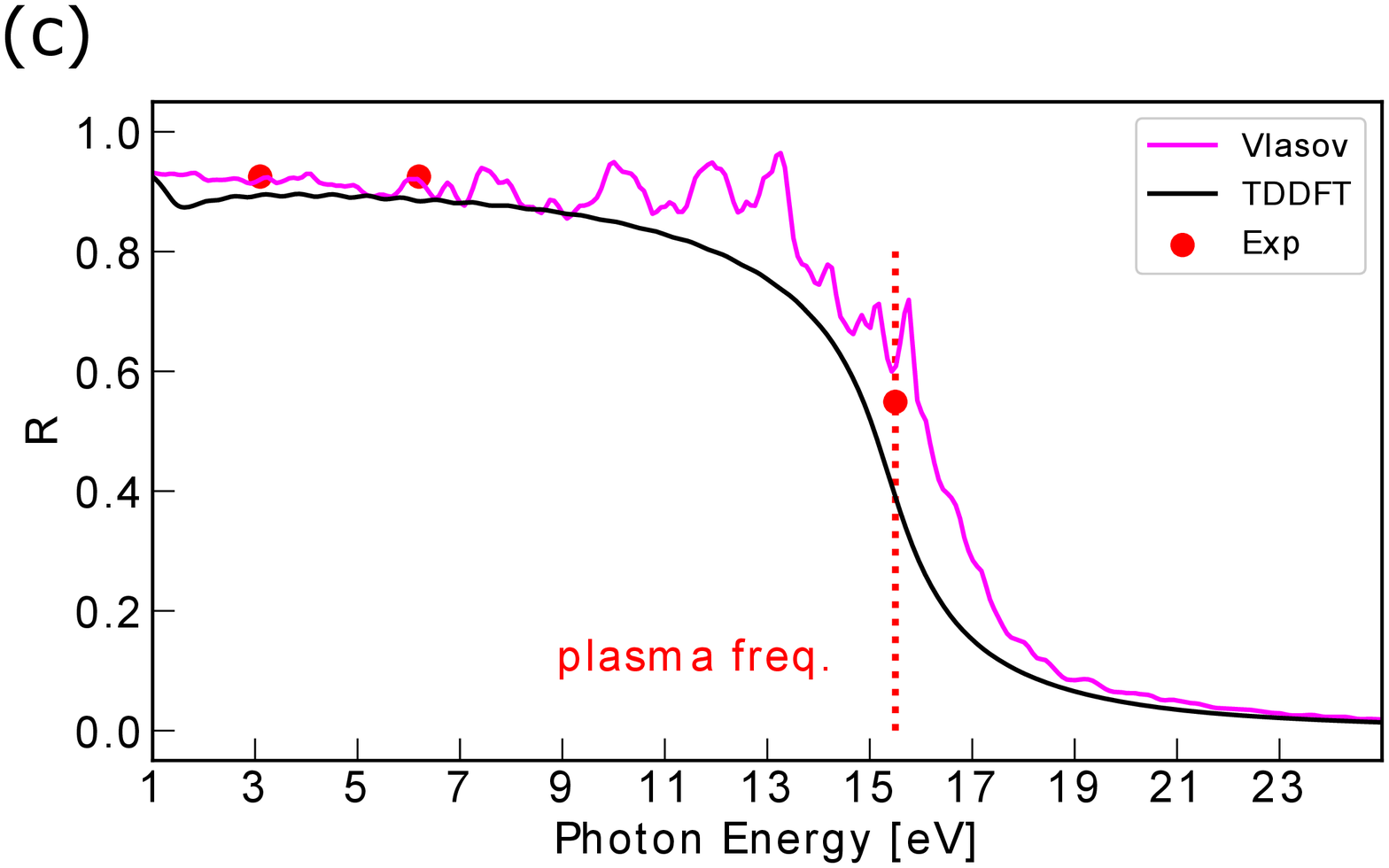}
  \caption[optical conductivity]{(a) Optical conductivity $\sigma(\omega)$, (b) refractive index $n$ and extinction coefficient $k$, (c) reflectivity $R(\omega)$, calculated with the Vlasov-LDA method and TDDFT as well as reported in experimental reference \cite{experiment}. The experimental values are plotted at 3.1 eV (400 nm), 6.2 eV (200 nm), and 15.5 eV (80 nm).}
  \label{LR}
 \end{figure}
 
\begin{figure}[tb]
  \centering
  \includegraphics[keepaspectratio,width=\hsize]{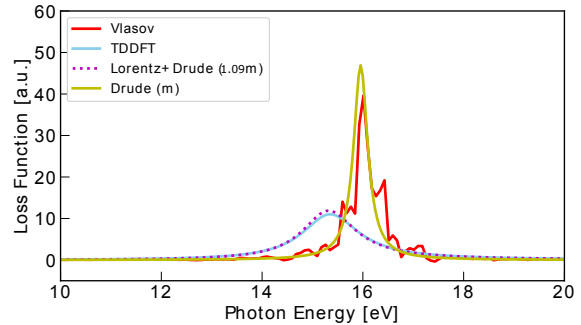}
  \caption[loss function]{Loss functions by TDDFT, Vlasov-LDA, combined Drude and Lorentz models ($m_{\mathrm{eff}}=1.09m$), and Drude model ($m_{\mathrm{eff}}=m$). The peak of the loss function gives plasma frequency.}
  \label{Loss}
 \end{figure}

Let us first discuss the complex optical conductivity, refractive index, extinction coefficient, and reflectivity as a function of photon energy.
Despite the simpleness of the Vlasov-LDA approach, its results excellently agree with the TDDFT results and experimental values (Fig.~\ref{LR}), especially above 2 eV photon energy.
The peak and dip around $1.5 \ \mathrm{eV}$ in the TDDFT results are due to interband absorption, which is not reproduced by the present Vlasov approach, since the latter takes only the single free-electron dispersion into account. Focusing on reflectivity behavior around the plasma frequency, one finds some differences between the two approaches. This difference would be contributions by the above-mentioned interband resonance and non-unity effective mass in TDDFT. 
We have confirmed it through the decomposition of the response obtained by TDDFT into Drude and Lorentz model components.
With the resonance energy set to 1.85 eV, the biggest oscillator around 1.5 eV \cite{DL2},
The estimated effective mass $m_{\mathrm{eff}}$ is $1.09m$, and the damping constant is $0.51 \ \mathrm{eV}^{-1}$, consistent with the values reported previously ($1.16m$ \cite{EffMass} and $0.80 \ \mathrm{eV}^{-1}$ \cite{DL2}, respectively).
The loss functions, $\mathrm{Im}\epsilon(\omega)^{-1}$, are shown in Fig.~\ref{Loss}.
The TDDFT result is excellently reproduced by the combined Drude and Lorentz contributions with $m_{\mathrm{eff}}=1.09m$.
Although Vlasov-LDA overestimates the plasma frequency compared to TDDFT, it agrees with the Drude model with $m_{\mathrm{eff}}=m$.

\subsection{\label{subsec:energy absorption} Energy absorption}

\begin{figure}[tb]
  \centering
  \includegraphics[keepaspectratio,width=\hsize]{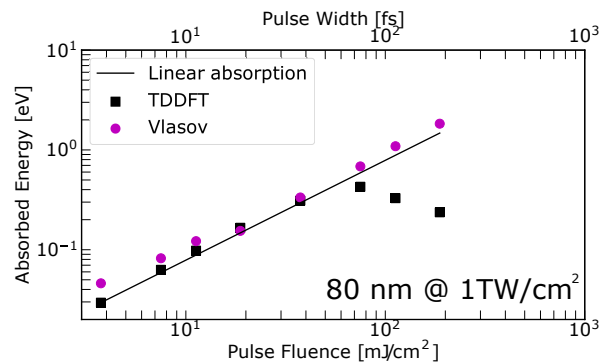}
  \caption[Pulse Duration dependence of electron energy absorption]{Calculated absorbed energy vs. pulse fluence or pulse width for a fixed intensity (1 ${\rm TW/cm}^2$). Pink circles: Vlasov-LDA, black squares: TDDFT, solid line: linear dependence passing through the square (TDDFT) for the $3.8 \ \mathrm{fs}$ pulse.
  }
  \label{energy-duration}
 \end{figure}

\begin{figure}[tb]
  \centering
  \includegraphics[keepaspectratio,width=\hsize]{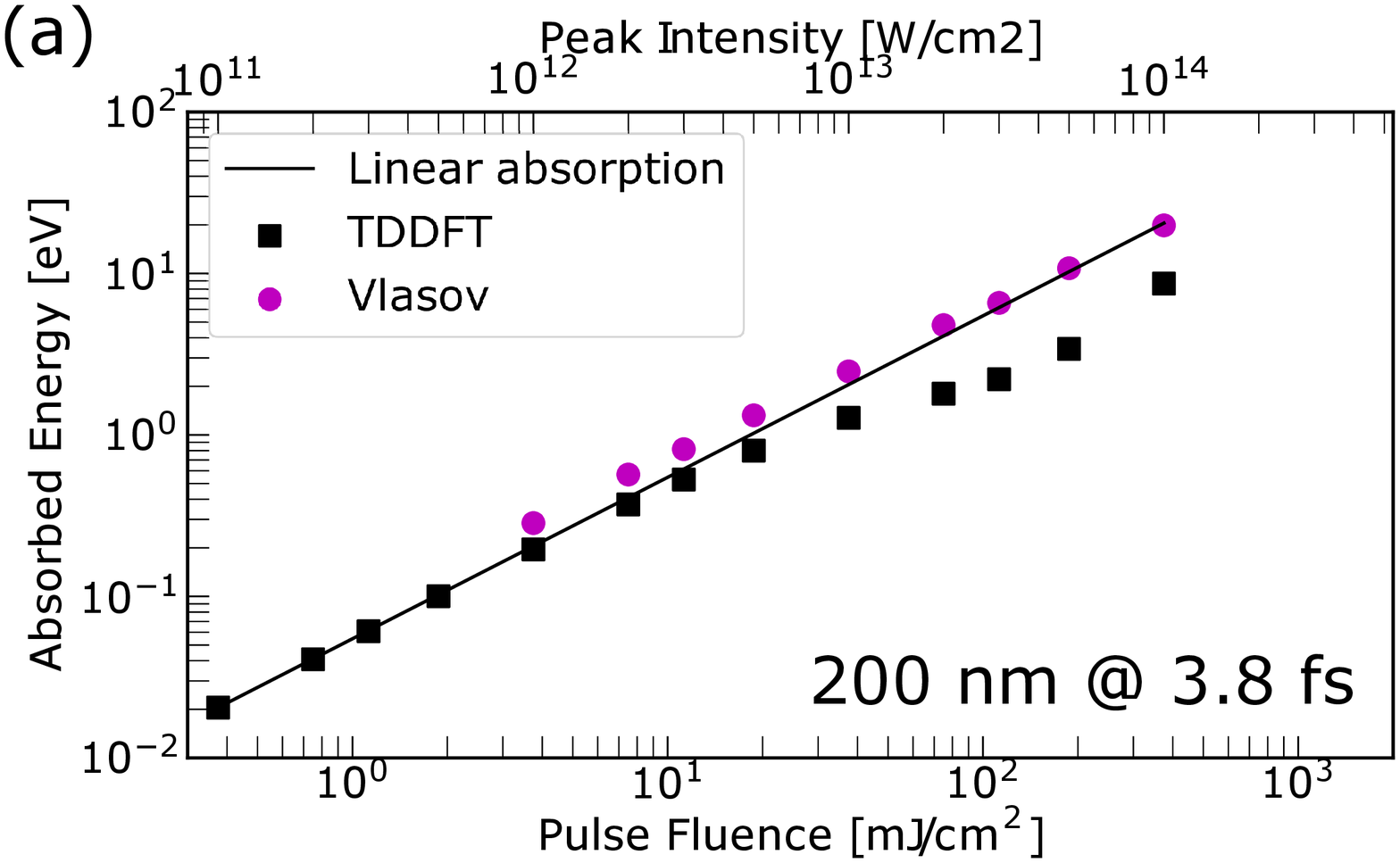}
  \includegraphics[keepaspectratio,width=\hsize]{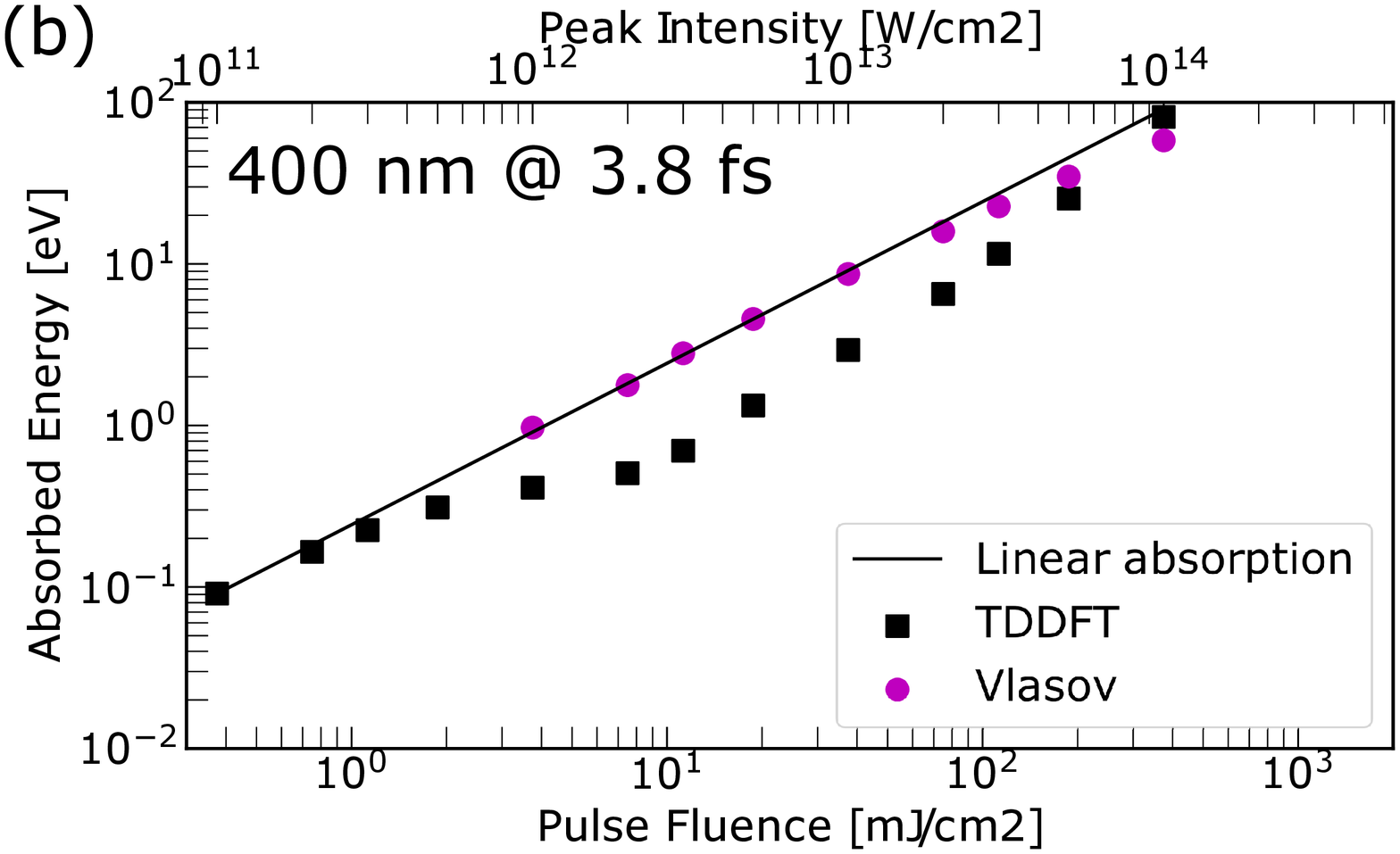}
  \caption[Laser intensity dependence of electron energy absorption]{Absorbed energy vs. pulse fluence or peak intensity for a fixed pulse width of 3.8 fs for the case of (a) 200 nm and (b) 400 nm wavelength. Pink circles: Vlasov-LDA, black squares: TDDFT, solid line: linear dependence passing through the square (TDDFT) for $10^{11} \ \mathrm{W/cm^2}$ intensity.}
  \label{energy-intensity}
 \end{figure}

Let us next investigate the energy absorption from the laser pulse.
We evaluate the energy absorption by the electrons as their energy increment by the pulse irradiation.
The energy is calculated as $\Delta E = H_{\mathrm{pp}}(t=\infty)-H_{\mathrm{pp}}(t=0)$ in the Vlasov-LDA simulation and as  $\expval{h_{\mathrm{KS}}(t=\infty)}-\expval{h_{\mathrm{KS}}(t=0)}$ in the TDDFT simulation.
We show the fluence dependence for the fixed intensity ($10^{12}\,{\rm W/cm}^2$) at 80 nm wavelength in Fig.~\ref{energy-duration} and that for the fixed pulse width of 3.8 fs at 200 and 400 nm wavelengths in Fig.~\ref{energy-intensity}. 

 \begin{figure}[tb]
  \centering
  \includegraphics[keepaspectratio,width=\hsize]{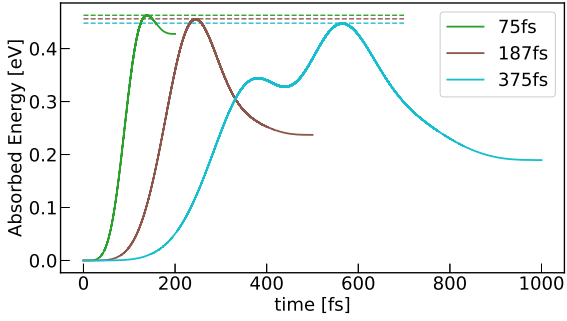}
  \caption[Pulse Duration dependence of electron energy absorption]{Solid lines: temporal evolution of the absorbed energy for three different pulse widths 75, 187, and 375 fs. Dashed lines: maximum values for each pulse width.
  }
  \label{rabi}
 \end{figure}
 
We see in Fig.~\ref{energy-duration} that both Vlasov-LDA and TDDFT results are linear in fluence and agree well with each other in the lower fluence region ($\lesssim 50 \ \mathrm{mJ/cm^2}$). 
On the other hand, the Vlasov-LDA does not reproduce the TDDFT results for the higher fluence, where the latter deviate from the linear behavior and even decrease with increasing fluence.
This difference is due to Rabi-like oscillation \cite{SA}, as confirmed in Fig.~\ref{rabi}, which shows the temporal evolution of absorbed energy for several pulse widths.
The maximum electron energy gain during the pulse, which are indicated by horizontal dashed lines, does not depend much on the pulse width, suggesting Rabi-like coherent oscillation.
Thus, there is an optimum pulse width for a fixed intensity in terms of energy absorption.

Figure \ref{energy-intensity} indicates that the energy absorption calculated by the Vlasov-LDA approach exhibit a linear dependence on fluence or pulse intensity in the whole range.
We can see a nonlinear behavior, on the other hand, in the TDDFT results in the higher fluence range ($\gtrsim 40 \ \mathrm{mJ/cm^2}$ for 200 nm and $\gtrsim 2 \ \mathrm{mJ/cm^2}$ for 400 nm.
This would be interpreted as saturable absorption as is widely observed in various materials \cite{SA2}.
We could not obtain the Vlasov-LDA results for the low fluence region ($\lesssim 2 \ \mathrm{mJ/cm^2}$) because of statistical error. 
This could be improved by increasing the total number of pseudo particles, in principle.
Nevertheless, the Vlasov-LDA results, if extrapolated to the low fluence region, appear to agree well with the TDDFT results.

\begin{figure}[tb]
  \centering
  \includegraphics[keepaspectratio,width=\hsize]{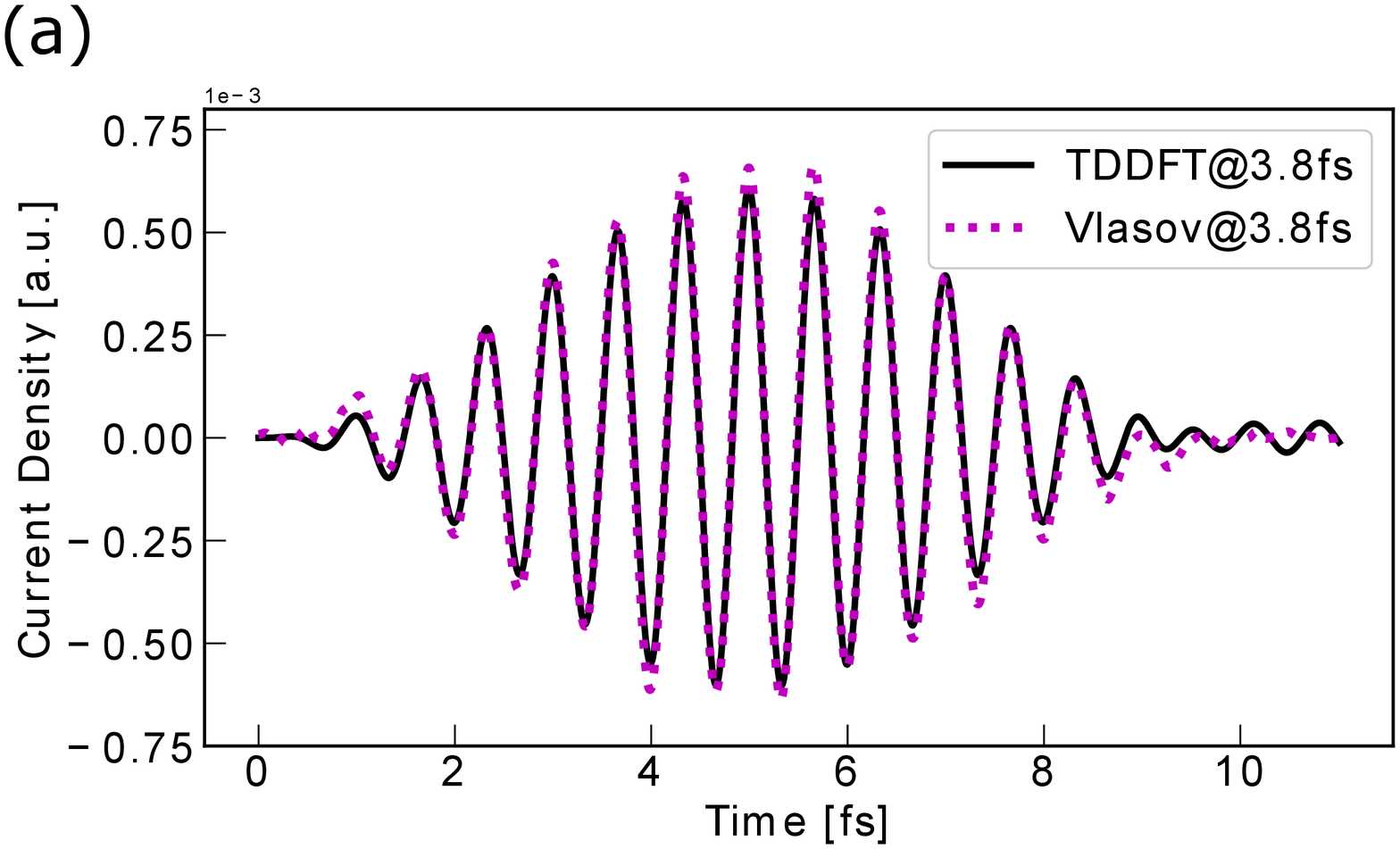}
  \includegraphics[keepaspectratio,width=\hsize]{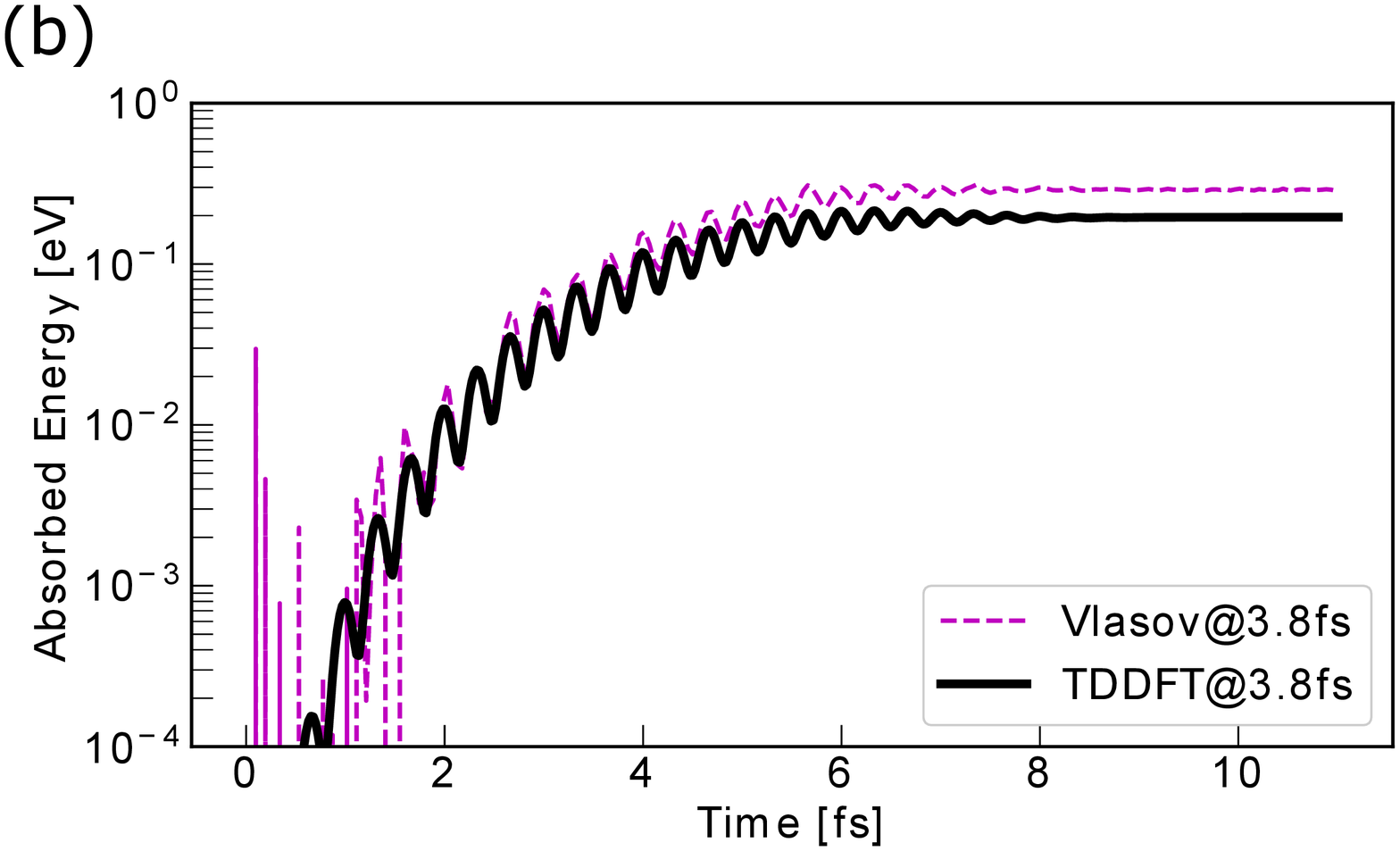}
  \caption[time-dependent quantities]{(a) Time-dependent current density $J(t)$ and (b) electron energy absorption $\Delta E(t)$. Pink dashed lines: Vlasov-LDA, black solid lines: TDDFT.}
  \label{td}
 \end{figure}

Figure \ref{td} shows the temporal evolution of the current density and absorbed energy for $10^{12}\,{\rm W/cm}^2$ peak intensity, $200 \ \mathrm{nm}$ wavelength, and 3.8 fs pulse width.
Again, overall, the Vlasov results excellently reproduce the TDDFT results.
In Fig.~\ref{td} (b), although energy fluctuation due to the pseudo-particle statistical error is seen at $<2\,{\rm fs}$, it becomes negligible at the end of the pulse.

Our code is partially parallelized using OpenMP and MPI. 
One of the most time-consuming parts is the Fourier transformation, which is computed by naive approach. Nevertheless, the computational time of the present Vlasov-LDA code is typically only 1/20 of that of TDDFT using the SALMON code. 
With more sophistication and parallelization, the efficiency of the Vlasov-LDA method will be further improved, which will be advantageous for applications such as parameter optimization in laser material processing.

\section{\label{sec:conclusions} Conclusions}

We have extended the Vlasov-LDA semi-classical approach and implemented it with the pseudo-particle method to periodic systems in order to compute the electron dynamics in solids, especially in metals, under ultrashort intense laser pulses.
The Vlasov equation can be regarded as the leading order of a semiclassical $\hbar$ expansion of the time-dependent Kohn-Sham equations.
The electronic distribution function is expressed by pseudo-particles, incorporating the periodic boundary condition.
They play the role of Lagrangian markers embedded randomly in the electron gas.
The initial distribution is calculated from the Thomas-Fermi model.

We have applied this approach to crystalline aluminum.
Although the method has only one adjustable parameter $d_r$, the calculated optical conductivity, refractive index, extinction coefficient, and reflectivity as well as energy absorption are overall in excellent agreement with the TDDFT and experimental results over a wide range of photon energy and fluence, demonstrating the capability of the present approach to accurately describe the dynamics of metallic conduction-band electrons.
On the other hand, the Vlasov results deviate from the TDDFT ones around 1.5 eV photon energy, where interband transitions are involved, and at the high-fluence region, where a Rabi-like oscillation takes place.


The next step will be to incorporate electron-electron collisions, the description of which is limited in TDDFT.
Vlasov-LDA is expected to provide valuable insights into complex laser-material processing if we further couple it with molecular dynamics, electromagnetic field analysis, and other continuum models.

\section*{\label{sec:acknowledgements} ACKNOWLEDGEMENTS}
We wish to express our gratitude to Kazuhiro Yabana for private discussions. This research was supported by MEXT Quantum Leap Flagship Program (MEXT Q-LEAP) Grant Number JPMXS0118067246. 
This research was also partially supported by JSPS KAKENHI Grant Number 20H05670, JST COI Grant Number JPMJCE1313, and JST CREST under Grant Number JPMJCR16N5. 
M.T. gratefully acknowledges support from the Graduate School of Engineering, The University of Tokyo, Graduate Student Special Incentives Program. M.T. also gratefully thanks support through crowd funding platform \it academist \rm by Misako Abe, Daigo Oue, Miho Otsuka, Yusaku Karibe, Ayano Sakai, Yushi Sakai, Shunsuke A. Sato, Ryosuke Shibato, Hitomi Suto, Tomoharu Sunouchi, Hideo Takahashi, and Yusuke Tokunaga. The numerical calculations are partially performed on supercomputers Oakbridge-CX, sekirei, ohtaka (the University of Tokyo), and SGI ICE X at Japan Atomic Energy Agency(JAEA). This research is partially supported by Initiative on Promotion of Supercomputing for Young or Women Researchers, Information Technology Center, The University of Tokyo.
\appendix
\def\thesection{\Alph{section}}

\end{document}